\documentclass[11pt, a4paper,fleqn]{article}
\usepackage[latin1]{inputenc}
\usepackage{amssymb}
\usepackage{amstext}
\usepackage{graphics}
\usepackage{graphicx}
\usepackage{authblk}
\usepackage{amsmath}
\usepackage{xcolor}

\include{defs}

\begin{document}

\title{Blinded sample size re-estimation in equivalence testing}

\author[1]{Ekkehard Glimm}
\author[2]{Lillian Yau}
\author[2]{Heike Woehling}

\affil[1]{Novartis Pharma AG, Basel, Switzerland}
\affil[2]{Sandoz Biopharmaceuticals, Hexal AG, Holzkirchen, Germany}

\maketitle

\begin{abstract}
This paper investigates type I error violations that occur when blinded sample size reviews are applied in equivalence testing. We give a derivation which explains why such violations are more pronounced in equivalence testing than in the case of superiority testing. In addition, the amount of type I error inflation is quantified by simulation as well as by some theoretical considerations. Non-negligible type I error violations arise when blinded interim re-assessments of sample sizes are performed particularly if sample sizes are small, but within the range of what is practically relevant.
\end{abstract}


\section{Introduction}
Discussions on sample sizes re-estimation (SSR) have appeared in statistical literature as early as the 1940\rq{s} \cite{Stein1945}. Since then a growing body of literature has emerged.
 For the two-sample $t$-test of a new treatment versus a control, methods based on estimators of the sample variance which can be calculated without unblinding the group affiliation of the observations available at an interim analysis have been introduced \cite{Kieser03, Friede06}. The simplest of these is the ``total variance estimate" which is the usual one-sample variance estimate calculated from treating all observations as if they were from only one treatment group. Although it has been shown that under this approach the type I error is not strictly controlled \cite{Proschan14, Lu16}, in the case of superiority testing, the type I error violations are extremely small and even then occur only in small samples. From an applications perspective, these investigations therefore reinforce the recommendations made by \cite{Kieser03} and \cite{Friede06}. However, in the case of non-inferiority (NI) testing, and thus also in equivalence (EQ) testing, this inflation can be larger \cite{Lu16, Friede03, Friede11, Friede13}. 
In this paper we take a closer look at the non-trivial type I error inflation in NI and EQ testing and the factors that impact it. We also provide practical recommendations to clinicians and data analysts who wish to implement a blinded SSR in their equivalence studies.

We base our discussion on a design with two stages and two treatment groups. Stage 1 consists of subjects whose data are used for estimating the total variance; stage 2 consists of subjects who are additionally recruited after the interim. In certain cases, no additional subjects are needed and stage 2 does not have any subjects. An important assumption is that beside the total sample size, other elements of the study design do not change after the interim.

We use the following notations in our discussions. For treatment group $j,\ j = \{1,\ 2\}$:
\begin{eqnarray}
\tilde{n}_j &=& \text{stage 1 sample size}\nonumber\\
m_j &=& \text{stage 2 sample size}\nonumber\\
n_j &=& \tilde{n}_j + m_j = \text{realized total sample size}\nonumber\\
n_{j, Min} &=& \text{planned minimum total sample size}\nonumber\\
n_{j, Max} &=& \text{planned maximum total sample size}\nonumber
\end{eqnarray}
We assume $n_{j, Max}\ge n_{j, Min}\ge \tilde{n}_j$.

In Section \ref{sec_typeI}, we derive a decomposition of the total variance observed at the time of blinded SSR to demonstrate how type I error inflation arises in NI and EQ tests. Section \ref{eqv_test} provides a brief introduction to the two one-sided tests (TOST) approach to EQ testings. Section \ref{sim_set} outlines the simulation settings and presents the results. Section \ref{sim_dis} discusses how different factors impact the type I error. In Section \ref{sec_summ} we summarize our findings.

\section{Type I error violation in non-inferiority testing}\label{sec_typeI}

A common strategy for EQ testing is the TOST approach \cite{Schuirmann87}. This uses two NI tests. For this reason we start by looking at the latter.

Assume that we want to test $H_0: \delta\geq \delta_0$ versus $H_A:\delta < \delta_0$ at level $\alpha$ on independent normal observations $y_{ij}$ from $N(\mu_j, \sigma^2)$ in two groups $j=1,2$ and $i=1,\ \dots\ n_j$. Let $\delta=\mu_1-\mu_2$, and $\bar{y}_j$ be the mean of the observations in group $j$. Without loss of generality, assume furthermore that $\delta_0 > 0$. We proceed with the ``total variance"
\begin{equation}\label{totvar}
\hat{\sigma}_T^2=\frac{Q}{\tilde{n}_{*}-1}=\frac{1}{\tilde{n}_{*} -1}\sum_{j=1}^2 \sum_{i=1}^{\tilde{n}_j} \left(y_{ij}-\bar{y}\right)^2,
\end{equation}
where $\tilde{n}_{*}=\tilde{n}_1+\tilde{n}_2$, and  $\bar{y}=\frac{\tilde{n}_1\bar{y}_1+\tilde{n}_2\bar{y}_2}{\tilde{n}_{*}}$ is the mean across the two groups.

To calculate $E(\hat{\sigma}_T^2)$, we decompose $Q$ into $Q_1$ and $Q_2$, where
\begin{eqnarray*}
Q_1 &=& \sum_{j=1}^2 \sum_{i=1}^{\tilde{n}_j} \left(y_{ij}-\bar{y}_j\right)^2\\
Q_2 &=& \sum_{j=1}^2 \tilde{n}_j \left(\bar{y}_j-\bar{y}\right)^2.
\end{eqnarray*}


It follows that $Q = Q_1 + Q_2$. By Cochran's theorem, $Q_1$ and $Q_2$ are stochastically independent with distributions $Q_1\sim\sigma^2 \chi^2(\tilde{n}_{*}-2)$ and $Q_2 \sim \sigma^2 \chi^2(1;\frac{\tilde{n}_1 \tilde{n}_2}{\tilde{n}_{*}}\frac{\delta^2}{\sigma^2})$, which is a non-central $\chi^2$ distribution. Hence,
$$E(Q)=E(Q_1)+E(Q_2)=(\tilde{n}_{*}-2)\sigma^2+\bigg(1+\frac{\tilde{n}_1 \tilde{n}_2}{\tilde{n}_{*}}\frac{\delta^2}{\sigma^2}\bigg)\sigma^2,$$

and
\begin{equation}\label{E_totvar}
E(\hat{\sigma}_T^2)=\frac{1}{\tilde{n}_*-1}E(Q)=\sigma^2\left(1+\frac{\tilde{n}_1 \tilde{n}_2}{\tilde{n}_{*}(\tilde{n}_{*}-1)}\frac{\delta^2}{\sigma^2}\right).\\
\end{equation}

\medskip

To illustrate how type I error inflation may arise, consider the following SSR rule:
\begin{enumerate}
\item If $\hat{\sigma}_T^2$ is large, i.e. $\hat{\sigma}_T^2 > c$ for some preselected threshold $c>0$, we recruit $m_j$, $j=1,2$, additional subjects.
\item If $\hat{\sigma}_T^2$ is small, i.e. $\hat{\sigma}_T^2 \leq c$, we do not recruit additional subjects, but stop and test right away with the $n =n_1+n_2= \tilde{n}_1+\tilde{n}_2$ subjects we have.
\end{enumerate}

If the true $\delta > 0$, we can see from Equation (\ref{E_totvar}) that a small value of $\hat{\sigma}_T^2$ provides evidence {\it against} $H_0$, because $E(\hat{\sigma}_T^2)$ does not only depend on the variance, but also on the squared true mean difference between the two groups. Hence, the rule above stops recruiting and uses the data obtained up to the interim analysis undiluted by additional observations if it happened to be in favor of $H_A$, but dilutes this evidence by recruiting more subjects if the data obtained up to the interim analysis shows evidence in favor of $H_0$. It is clear that this must inflate the type I error of the entire strategy. The same is true for other blinded sample size re-estimation rules where the final total sample size is an increasing function of $\hat{\sigma}_T^2$. On the other hand, if the sample size re-estimation rule were a decreasing function of $\hat{\sigma}_T^2$, then we would have $\alpha$ deflation.

Two remarks are important. Firstly, from Equation (\ref{E_totvar}) we can see that the inflation is primarily associated with the shifted hypothesis $H_0: \delta \geq \delta_0 > 0$. If $\delta_0=0$, then at the point of $\delta=\delta_0$, $\hat{\sigma}_T^2$ is an unbiased estimate of $\sigma^2$. Regarding type I error control, what matters is the behavior of the test at $\delta=\delta_0$. Hence, the issue investigated here does not occur with superiority testing where $H_0: \delta = \delta_0 = 0$.  Intuitively, a large $\hat{\sigma}_T^2$ does not provide evidence in favor of, or against, the superiority hypothesis, because a large observed $\hat{\sigma}_T^2$ can equally likely arise from $\delta$ or $-\delta$ being true. With the shifted $\delta=\delta_0>0$, however, a large value of $\hat{\sigma}_T^2$ will much more likely occur with a value of $\bar{y}_1-\bar{y}_2 > \delta$ than with $\bar{y}_1-\bar{y}_2 < -\delta$. We note in passing that simply subtracting a constant, i.e. a hypothetical true difference, from the total variance cannot influence this issue.

Secondly, the statements we have made here are qualitative: There will be a type I error inflation, but its magnitude depends on many things:
the {\em effect size} $\delta_0/\sigma$, the stage 1 sample size, the range of possible values of the stage 2 sample size, the actual small sample properties of a test using a test statistic which is only asymptotically Normal, and the like. Equivalence testing with TOST is a case in point.  If the equivalence margin is symmetric (i.e. $\delta_0=\delta_{up}=-\delta_{low}$), then exactly the same mechanisms are at work at both the upper and the lower end of the equivalence range, and both one-sided level-$\alpha$-tests are subject to type I error inflation with the sample size review strategy described above. The actual probability of rejection in the TOST, however, is also subject to the inherent conservatism of the TOST strategy which arises from the fact that there is a positive probability of not rejecting $H_{01}: \delta \le -\delta_0$ when in fact $H_{02}: \delta \ge \delta_0$ is true. This conservatism of course remains and may counteract against the liberality caused by the phenomenon we discussed.

An analytical calculation of the exact type I error rate is possible, however, given the many factors at play, the derivation is often algebraically consuming. In Appendix \ref{app_num}, we discuss the quantification of the type I error inflation in a specific case. For general discussion, we use simulation studies.

\section{Equivalence testing and TOST}\label{eqv_test}

In the TOST approach to equivalence testing, two hypotheses are tested simultaneously:
\begin{equation*}
\text{H}_{01}: \delta \le \delta_{low}\ \text{vs.}\ \text{H}_{A1}: \delta > \delta_{low},\text{ and}\\ 
\end{equation*}
\begin{equation*}
\text{H}_{02}: \delta \ge \delta_{up}\ \text{vs.}\ \text{H}_{A2}: \delta < \delta_{up}, \\ 
\end{equation*}
where $\delta_{low}$ and $\delta_{up}$ are the lower and the upper {\em equivalence margins}. To claim equivalence, both null hypotheses must be rejected simultaneously at level $\alpha$. Using the same notations as in Section 2, and with the given {\em reference range} $(\delta_{low},\ \delta_{up})$, we reject the null hypothesis of non-equivalence $H_0 = H_{01}\cup H_{02}$ if
\begin{equation*}
\sqrt{\frac{n_1n_2}{n_1+n_2}}\cdot\frac{d-\delta_{low}}{s}> t_{1-\alpha}(n_1+n_2-2),
\end{equation*}
and
\begin{equation*}
\sqrt{\frac{n_1n_2}{n_1+n_2}}\cdot\frac{d-\delta_{up}}{s}< t_{\alpha}(n_1+n_2-2),
\end{equation*}
where $d=\bar{y}_1-\bar{y}_2$ and
\begin{equation*}
s^2 = \frac{1}{n_1+n_2-2}\sum_{i=1}^2\sum_{j=1}^{n_i}(y_{ij}-\bar{y_i})^2. 
\end{equation*}

This decision rule can also be described via a $(1-2\alpha)$ level confidence interval (CI) for  $\delta$. If this CI is completely contained within $(\delta_{low},\ \delta_{up})$, then equivalence between the two groups is declared at level $\alpha$.

\begin{figure}[ht!]
  \centering
  \includegraphics[width=1\textwidth]{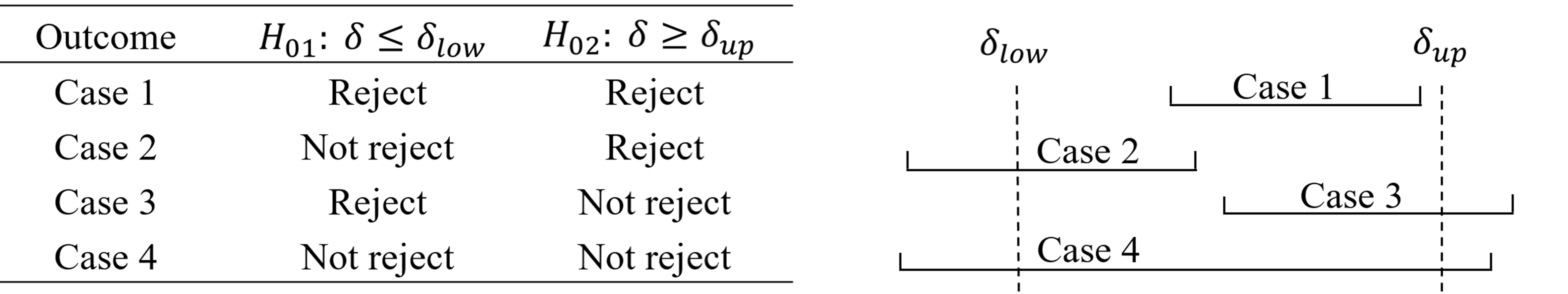}
 \caption{Four possible outcomes of a TOST}\label{fourCases}
\end{figure}

In Figure \ref{fourCases} we distinguish four possible outcomes with regard to the rejection of the two null hypotheses, and illustrates them in terms of the CI for $\delta$.  Per these definitions, Case 1 is where equivalence is established.

Consider an EQ test using TOST where $H_{02}$ is true. Case 1 represents a false rejection and establishes an equivalence incorrectly, i.e. a type I error. Although Case 2 is the least desirable outcome, it makes TOST conservative: it falsely rejects $H_{02}$, but also falsely does not reject $H_{01}$, thus comes to the correct conclusion that the equivalence cannot be established, {\em even though the reasoning is wrong}. Case 3 is the most desired outcome since it correctly supports non-rejection of $H_{02}$ but rejection of $H_{01}$. Case 4 does not reject either null hypothesis, therefore, reaches the correct decision of not rejecting $H_{02}$.

For NI testing where $H_{02}$ is true, Case 1 and Case 2 together represent false rejection of the null hypothesis.

\section{Simulation studies}\label{sim_set}

We investigate the impact of stage 1 sample size ($\tilde{n}_j$), stage 2 sample size ($m_j$), and effect size on type I error inflation in a blinded SSR for EQ testing using simulation studies. Without loss of generality, we set $\sigma$ to 1, and assume symmetric equivalence margins where $\delta_{up}=\delta_0=-\delta_{low}$; hence, in subsequent discussions $\delta_0$ and effect size $\delta_0/\sigma$ are numerically the same.

The sample sizes for the two groups are assumed the same at every stage, so the subscript $j$ is dropped in the discussion. We follow this strategy: before the start of the study, define $\tilde{n}$, $n_{Min}$, and  $n_{Max}$ where $\tilde{n}\leq n_{Min}\leq n_{Max}$. A blinded sample size re-assessment is conducted when data from $\tilde{n}$ subjects per group become available.

\subsection{The SSR rule}\label{ssr_rule}
The total sample size needed in each group, denoted $\hat{N}$, in order to reject $H_{01}$ and $H_{02}$ for a desired power $1-\beta$ is calculated based on $\hat{\sigma}_T^2$. $\hat{N}$ is compared to $n_{Min}$ and $n_{Max}$ such that
\begin{itemize}
\item[(1)] if $\hat{N} \le n_{Min}$, then additional $m = n_{Min} - \tilde{n}$ subjects are recruited per group in the second stage;
\item[(2)] if $\hat{N} > n_{Min} $, then additional $m = \text{min}(\hat{N} , n_{Max}) - \tilde{n}$ subjects are recruited in the second stage.
\end{itemize}
Hypotheses $H_{01}$ and $H_{02}$ are tested with $n = \tilde{n}+m$ total subjects per group at the end of the study. 

\subsection{Simulation settings}

All of the simulations use $\alpha=5\%$ and $\beta=10\%$. Table \ref{table1} summarizes the other conditions:

\begin{table}[ht!]
\caption{Simulation settings\label{table1}}
\centering
\begin{tabular}{ll}
\hline
Parameter & Values\\ \hline
$\tilde{n}$ & 10, 15, 20, 25, 30, 40, 50, 60\\
$n_{min} / \tilde{n}$ & 1, 1.2, 1.4, 1.6, 1.8, 2\\
$n_{max} / \tilde{n}$ & 2, 2.5, 3, 3.5, 4, infinity\\
$\delta_0/\sigma = \delta_0$ & 0.05 to 1.5 at 0.05 interval\\ \hline
\end{tabular}
\end{table}

Take $\tilde{n}=20$,  $n_{Min} / \tilde{n}=1.4$ (or equivalently $n_{Min} = 20\times 1.4 = 28$), and $n_{Max} / \tilde{n} =2$ (or $n_{Max} = 20\times2 = 40$) as an example. Our strategy says that the interim analysis is to be conducted when data from 20 subjects per group become available. Based on the total variance estimate $\hat{\sigma}_T^2$, if the total sample size $\hat{N}$ is determined to be no more than 28 subjects per group, then additional $m=8$ subjects per group will be recruited in the second stage, making it total of $n=n_{Min}=28$. On the other hand, if the re-assessed sample size $\hat{N}$ turned out to be larger than 28, then $m = \text{min}(\hat{N}, 40)-20$ subjects per group are to be recruited for the second stage.

One million simulations are performed for each combination of the parameter values in Table \ref{table1}. Normally distributed data are generated for the two groups with a mean difference of $\delta_0$. This corresponds to a true $H_{02}$. In each simulated sample $t$-test is conducted between the two groups. The percentages of each of the four cases (as illustrated in Figure \ref{fourCases}) are summarized.

\subsection{A note on sample size calculation}
For each simulated sample, the total sample size $\hat{N}$ is calculated from the stage 1 data by assuming a common population variance $\sigma^2$ estimated by $\hat{\sigma}_T^2$. Ideally $\hat{N}$ should be calculated using non-central $t$-distribution as outlined in \cite{Schuirmann87}. However, this would require iterative approximation of degrees of freedom for each single simulation. Furthermore, we focus on type I error inflation which ultimately depends on the sample size formula only via $\hat{N}$. For this purpose, it is essentially irrelevant if the $t$- or the simpler $z$-formula is used. In both cases, type I error inflations follow identical patterns. Therefore, we use the usual sample size formula based on the asymptotic normal distribution:
\begin{equation}\label{eqSS}
\hat{N} = \frac{2(Z_{1-\beta/2} + Z_{1-\alpha})^2}{((\delta_0-D)/\hat{\sigma}_T)^2},
\end{equation}
assuming 1:1 randomization ratio.  Since we are blinded at interim analysis, we do not have an estimate for $\delta$; a value $D$ must be assumed. In NI and EQ testing, it is common to assume that there is no difference between groups when doing power and sample size calculations. We follow this practice and assume $D=0$.

Although Equation (\ref{eqSS}) tends to give smaller sample sizes, its dependency on $\delta_0/\sigma_T$ is the same as its $t$-distribution based counterpart. 
Other components such as $\tilde{n},\ m,\ n_{Min},\ n_{Max}$, etc., only impact the amount of the inflation. As Appendix \ref{app_num} shows, a different sample size determination rule may dictate a completely different pattern of $\alpha$ behavior. Since Equation (\ref{eqSS}) and its $t$-distribution counterpart are implemented in many statistical software for power and sample size calculations, for example NCSS PASS or R package PowerTOST \cite{Hintze11, Labes18}, we make this our main objective of investigation.

\subsection{Results}\label{sim_res}
\begin{figure}[h]
\centering
\includegraphics[width=1\textwidth]{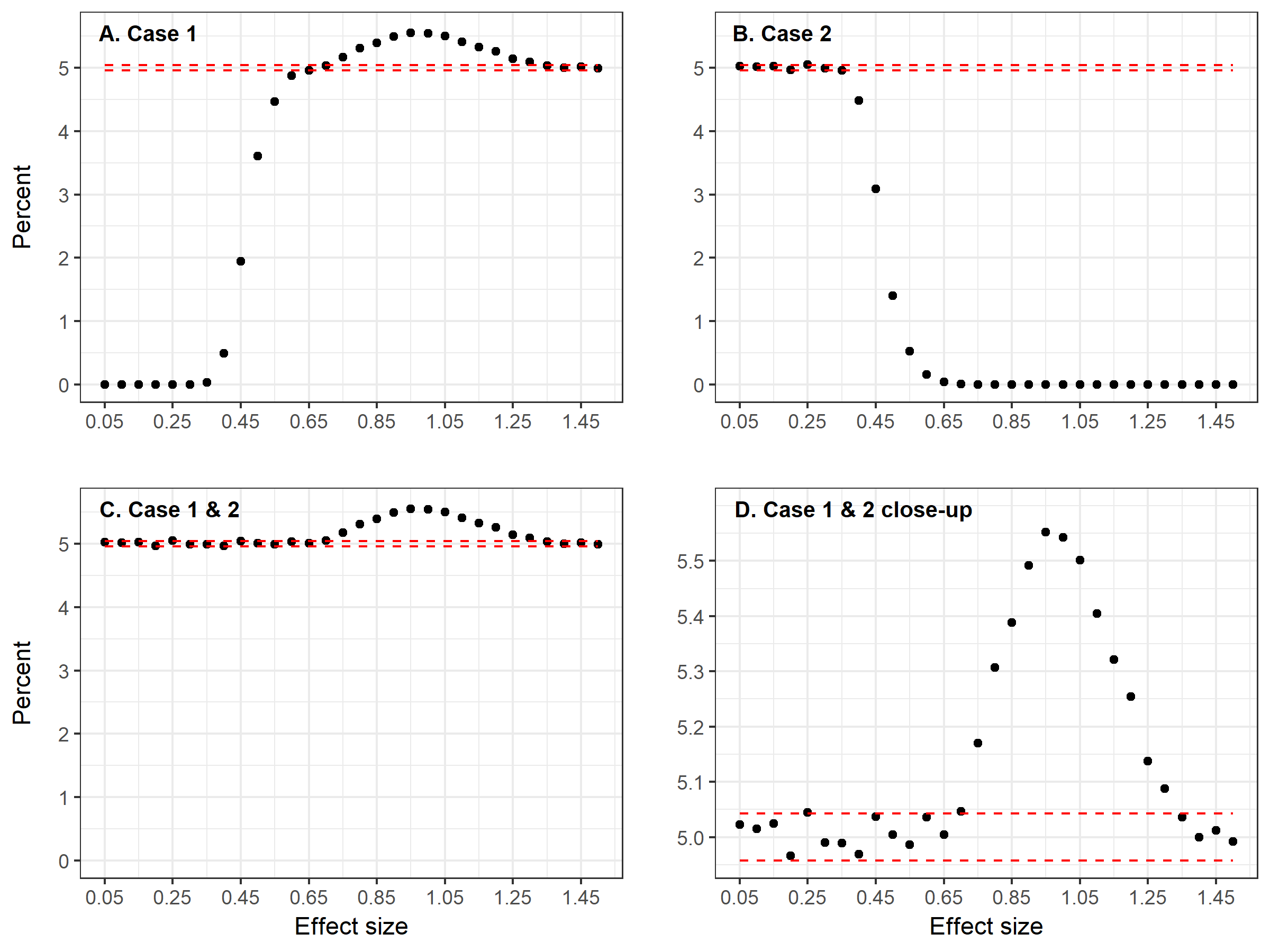}
\caption{Percent Case 1 or Case 2}\label{nInt15Min18Max30}
\end{figure}

We first present a benchmark example. Figure \ref{nInt15Min18Max30} shows the observed type I error rate for NI and EQ testings for the setting where $\tilde{n}= 15$, $n_{Min}=18$, and $n_{Max}=30$. What this rule says is that we conduct the interim when there are 15 subjects in each group; if the observed total variance $\hat{\sigma}_T^2$ is small enough such that we will only need up to a total of 18 subjects per group in order to reject $H_0$ with 90\% power at 5\% nominal $\alpha$ level, then we will recruit 3 additional subjects in each group; otherwise, we will recruit however many subjects needed but not exceeding 30.

In Figure \ref{nInt15Min18Max30} there are two red dashed lines on each panel representing the 95\% CI around 5\% in one million simulations. Any values within these two dashed lines can be attributed to random variation. In Panel A, percentages of Case 1 are plotted against the effect size. Recall a Case 1 is where the CI for the difference between the two groups is completely contained within the equivalence margins. Its probability under $H_0$ is the type I error rate for an EQ testing. When effect size is at or below 0.30, the standard deviation $\sigma$, which is set to be 1 for all the simulations, is much too large for $\delta_0$. The interim total variance $\hat{\sigma}_T^2$, serving as an approximation for $\sigma^2$, will tend to be very large in this area as well. Therefore, even with the largest sample size allowed according to our SSR rule, nearly no CI for the difference between the two groups are narrow enough to be within $(-\delta_0, \delta_0)$. Hence, no type I error is committed under the EQ hypothesis.

However, this is not to say that $H_{02}: \delta \ge \delta_0$ is correctly not rejected. Panel B of Figure \ref{nInt15Min18Max30} shows the percent of Case 2 against effect size. Recall a Case 2 is where the lower end of the confidence interval is below the lower margin. The probability of false rejection of $H_{02}$ starts out at 5\% and drops below that at effect size equals to 0.30 for exactly the reason that $\sigma$ is too large relative to $\delta_0$ in this area. If $H_{02}$ is indeed falsely rejected, the chances are higher that they are Case 2 rather than Case 1.

As $\delta_0$ becomes larger the CI for the group difference is becoming narrower given the same SSR rule. This in turn shows in Panel A such that the percentage of Case 1 is increasing sharply between effect size 0.30 and 0.5; at the same time, percentage of Case 2 starts to decrease, and reaching 0\% at 0.55 just when percent Case 1 reaches 5\%. If we look at Panel C, we see the two cases add up nicely to about 5\% or exceeding that where there is inflation. In fact, Panel C (or Panel D for a close-up) are the observed $\alpha$ in a NI test against $H_{02}$ with the given SSR rule. For both NI and EQ testing, the $\alpha$ inflation happens between effect size 0.70 and 1.25 with a peak at 0.95.

\begin{figure}[h]
\includegraphics[width=1\textwidth]{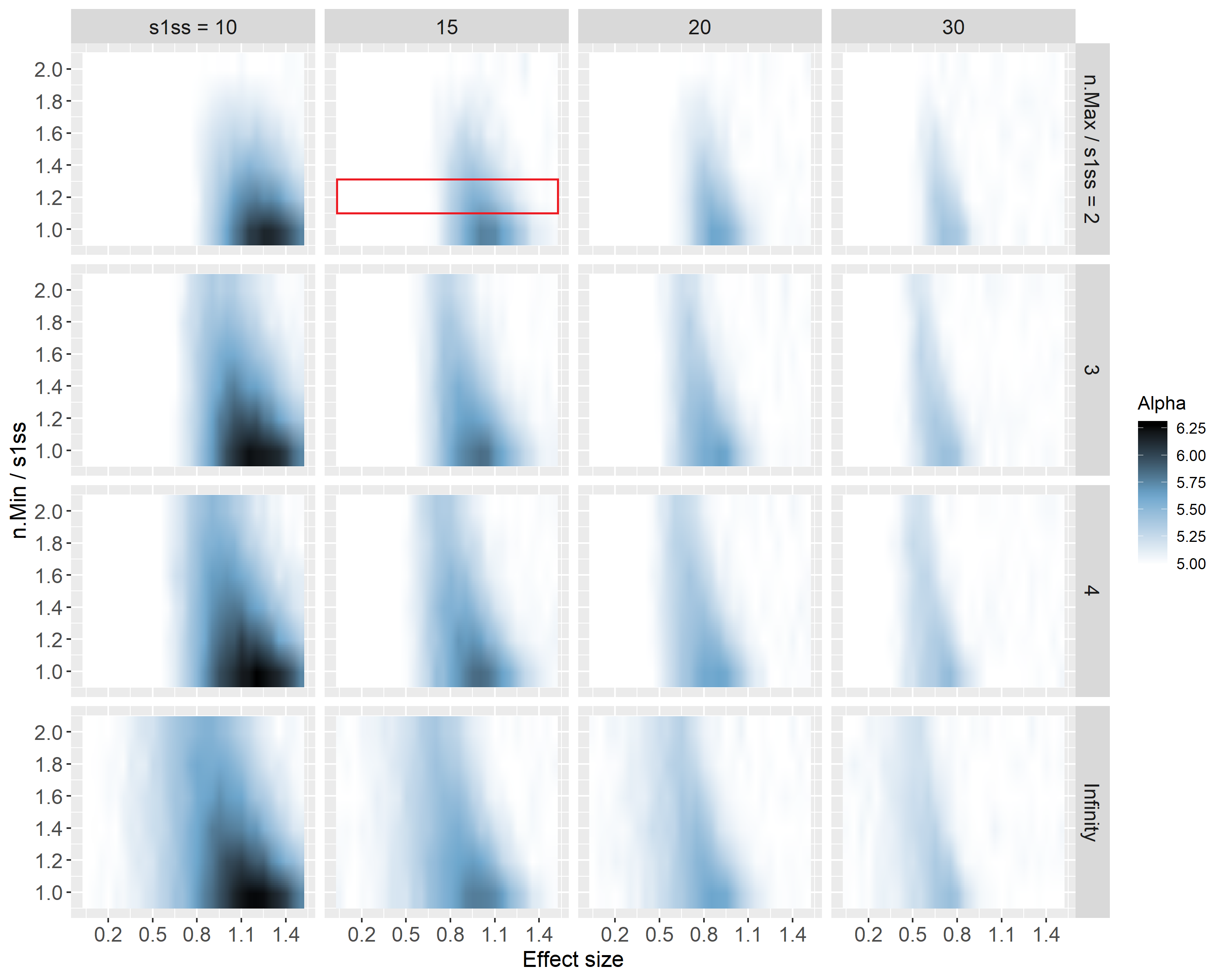}
\caption{$\alpha$ and inflation for stage 1 sample sizes $\tilde{n}$ (or s1ss) = 10, 15, 20, 30.}\label{heat4by4}
\end{figure}

The patterns in Figure \ref{nInt15Min18Max30} are consistent in all of the settings we examined. The difference is in where the peak $\alpha$ occurs, and how much it inflates to. Figure \ref{heat4by4} summarizes the percent of Case 1 in form of heatmaps where a darker shade indicates a larger $\alpha$ value. In addition, we see how $\tilde{n}$, $n_{Min}$, and $n_{Max}$ influence $\alpha$ inflation from these plots. Four stage 1 sample sizes, $\tilde{n}=10,\ 15,\ 20,\ \text{and}\ 30$, are selected for display. They are in columns from left to right. The rows in Figure \ref{heat4by4} represent $n_{Max}$ in terms of the ratio $n_{Max}/\tilde{n}$. The selected values are $n_{Max}/\tilde{n}=2,\ 3,\ 4,$ and infinity. Within each of the 16 heatmaps, the $x$-axis represents the effect size from 0.05 to 1.5; the $y$-axis represents the $n_{Min}$ in terms of the ratio $n_{Min}/\tilde{n}$. Note that in our simulations we have chosen discrete $n_{Min}/\tilde{n}$, and the y-axis is not truly continuous. The benchmark example we have shown in Figure \ref{nInt15Min18Max30} is highlighted with a red rectangle.

The graphs show that inflations of type I errors occur at $\delta_0/\sigma$ between 0.8 and 1.5, and that they are more severe with small stage 1 sample sizes and when $n_{Min}=\tilde{n}$.

\section{Discussion}\label{sim_dis}

\subsection{Stage 1 sample size and $\alpha$ inflation}\label{sec_tilde_n_a}

In Figure \ref{heat4by4} we see a consistent decrease of $\alpha$ inflation as $\tilde{n}$ increases. This is related to the precision in the estimate of $Var(\hat{\sigma}_T^2)$. Given Equation (\ref{totvar}), we derive
\begin{equation}\label{varHatSigT}
Var(\hat{\sigma}_T^2) = \frac{2\sigma^2}{(2\tilde{n}-1)^2}\left[\tilde{n}(2+\delta^2)-1\right] =\mathcal O\left(\frac{\delta^2}{\tilde{n}}\right).
\end{equation}
With fixed $\delta$, $Var(\hat{\sigma}_T^2)$ decreases as $\tilde{n}$ increases. The sample size formula in Equation (\ref{eqSS}) which is investigated in the simulations depends on the stage 1 sample size only via $\hat{\sigma}_T^2$. The more variation we see in $\hat{\sigma}_T^2$ the more variable second stage sample sizes will be. In turn this variability in stage 2 sample sizes induces $\alpha$ inflation.

To have a closer look, we investigate how the percentage of rejections depends on $\hat{\sigma}_T^2$ for an exemplary case with  $\delta_0/\sigma=1$ between a fixed design and 2-stage design with SSR. Results are from $10^5$ simulations assuming $H_{02}: \delta=\delta_0$ is true. First, the responses from a total of 30 subjects, 15 per group, are generated. For the fixed design, ordinary $t$-tests are conducted from these 30 subjects. For the 2-stage design with SSR, the total variance $\hat{\sigma}_T^2$ is calculated from the 30 subjects, and a stage 2 sample size $m$, where $0\le m < \infty$, is determined according to Equation (\ref{eqSS}) with $\alpha=5\%$ and $\beta=10\%$. Data from the two stages are then combined to conduct the $t$-test across both stages.

\begin{figure}[h]
\centering
\includegraphics[width=1\textwidth]{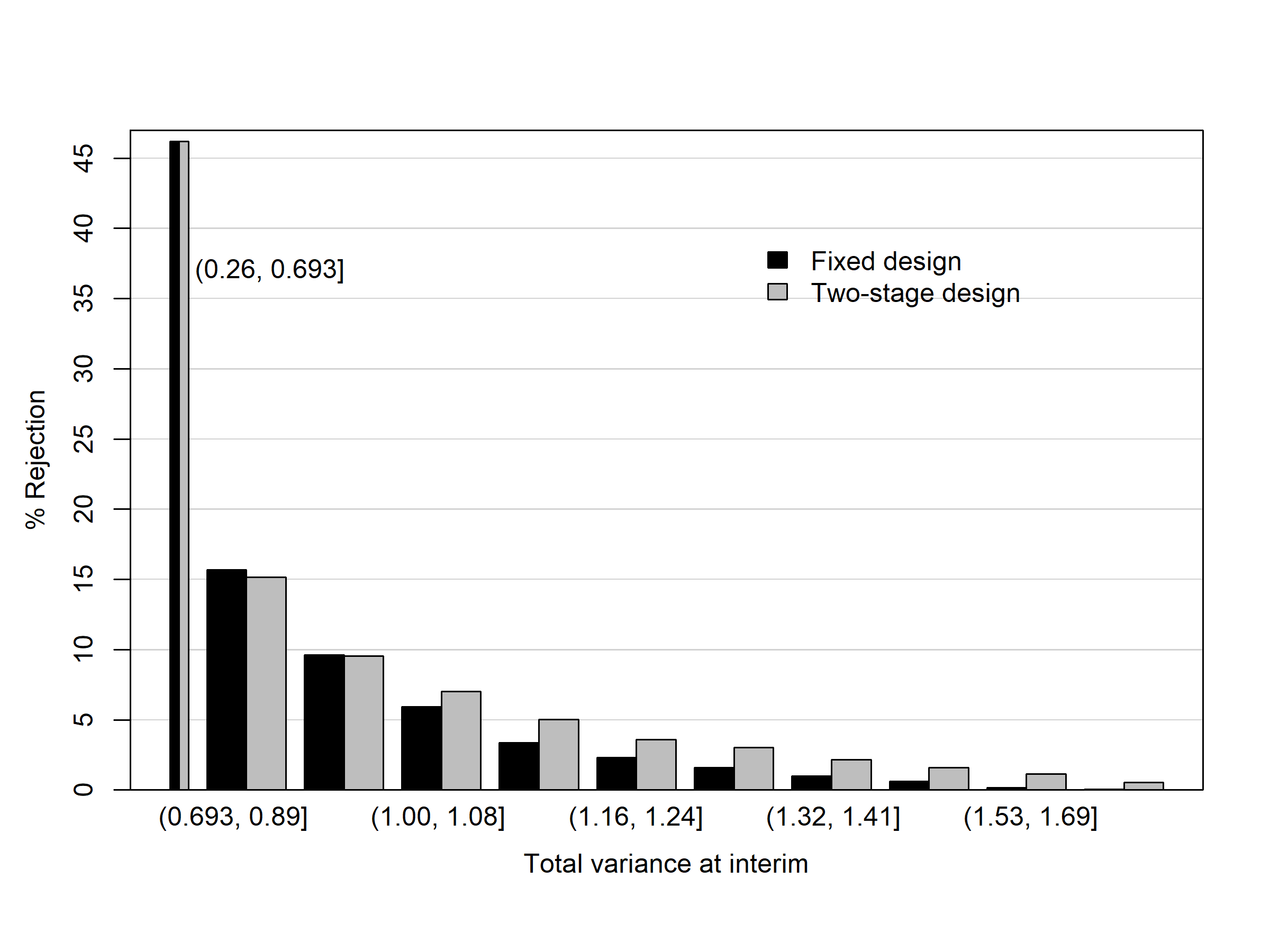}
\caption{Percent false rejections in intervals of total variance $\hat{\sigma}_T^2$, $\delta_0/\sigma=1$.}\label{varTalpha}
\end{figure}

Figure \ref{varTalpha} shows the percent of rejections of $H_{02}$ is related to the distribution of values of $\hat{\sigma}_T^2$ for both the fixed design (black bars) and the 2-stage design (gray bars). The height of the bars shows the percentage of rejections which occur in intervals of observed $\hat{\sigma}_T^2$ values. The first of these intervals contains all values of $\hat{\sigma}_T^2$ which are so small that $m=0$. In this example, this requires that $\hat{\sigma}_T^2\le 0.693$, and includes 2.3\% of all simulations. When this happens, the fixed and the 2-stage design are the same. $46\%$ of the $2.3\%$ where $\hat{\sigma}_T^2\le 0.693$ (i.e. 1.06\% of all cases) in the simulation are rejections of $H_{02}$. The remaining cases (where $\hat{\sigma}_T^2>0.693$ and hence $m>0$) are binned into 9 equally sized intervals, i.e. each interval contains 9,771 simulation results.

The plot supports the results from Section \ref{sec_typeI} which show that small values of $\sigma_T^2$ are providing information against $H_{02}$. As $\sigma_T^2$ gets larger, the percent rejections by the fixed design are decreasing towards 0: the weighted average across all intervals is of course $\alpha=5\%$. For the 2-stage design, the rejection rate is the same as the fixed design when $\hat{\sigma}_T^2\le 0.693$ (because $m=0$ there), but as $\hat{\sigma}_T^2$ increases, they diverge from each other. While the second (from the left) gray bar is even lower than its black counterpart, from the 4th pair onwards the 2-stage rejection rates are higher. This reflects the fact that additional stage 2 data helps to overcome the dampening effect of large $\sigma_T^2$ on rejection probability. Overall, the weighted average of the gray bars becomes $5.83\%$ demonstrating the type I error inflation.

The graph also shows that while general tendencies are predictable---such that rejection rates must decrease; the two sets of rates diverge as we move to the right; the gray bars on average are higher than the black bars especially on the right of the plot, etc.---some peculiarities of the specific sample size formula in Equation (\ref{eqSS}) also play a role. In this case, for example, the largest total sample size requested by any of the simulation runs is 74. This shows that, although we have not put any \lq\lq{}formal\rq\rq{} upper limit on stage 2 sample size, the formula is such that stage 2 sample sizes which completely dominate the stage 1 data simply do not occur. If the sample size formula were such that huge stage 2 sample sizes would occur frequently enough, the red bars would approach a height of $5\%$ as we move to the right in the plot. The fact that the 2-stage design has a slightly lower rejection rate in the second lowest interval of $\sigma_T^2$ can be explained as such. 

In the case when $\tilde{n}$ is large, the observed range for $\hat{\sigma}_T^2$ is much smaller. The behavior of the observed $\alpha$ values  resembles a fixed design more closely. The inflation of $\alpha$ is less severe.

As this reasoning shows, the variability of $\hat{\sigma}_T^2$ in the simulations is a good indicator of the SSR method's ability to react to different stage 1 data situations in favor of rejection of $H_0$. In contrast, when $\hat{\sigma}_T^2$ is very precise, the total (and the second stage) sample size is almost always the same from simulation to simulation, and we are led to always need the same overall sample size. This resembles a fixed stage design where the total sample size is predetermined and a mid-way look of the data does not change any element of the study design, hence, having no impact on $\alpha$.

Therefore, one obvious remedy in reducing $\alpha$ inflation is to increase the stage 1 sample size. However, in practice stage 1 sample size has to be reasonably small so that operationally there is enough time for an interim analysis to be carried out. Alternatively, we could reduce the variability of the stage 2 sample size by imposing a $n_{Min}$ or a $n_{Max}$ on the overall sample size. Of course, this is against the original motivation for the sample size review whose intent is to be able to modify the final actual sample size as freely as possible.


\subsection{Effect size and $\alpha$ inflation}
From Equations (\ref{totvar}) and (\ref{eqSS}), with $\sigma=1$, $D=0$, and under $H_{02}: \delta=\delta_0$, we derive the expected value of the total sample size for each group
\begin{equation}\label{eHatN}
E(\hat{N})=2(Z_{1-\beta/2} + Z_{1-\alpha})^2\left(\frac{1}{\delta_0^2}+\frac{\tilde{n}}{4\tilde{n}-2}\right),
\end{equation}
and its standard deviation
\begin{equation}\label{sdHatN}
SD(\hat{N}) = 2(Z_{1-\beta/2} + Z_{1-\alpha})^2\times\frac{SD(\hat{\sigma}_T^2)}{\delta_0^2} = \mathcal O\left(\frac{1}{\delta_0}\cdot\frac{1}{\sqrt{\tilde{n}}}\right).
\end{equation}
Although both $E(\hat{N})$ and $SD(\hat{N})$ approach infinity as $\delta_0$ goes to 0, the former approaches at a higher rate since
$$
\lim_{\delta_0\rightarrow0} \frac{E(\hat{N})}{SD(\hat{N})}=\frac{1}{\delta_0}=\infty.
$$
For small $\delta_0$, stage 2 sample size will always overwhelm stage 1 sample size if there is no restriction on the overall sample size.
In our simulations, when $\delta_0=0.05$ the average total sample size required to reject the null hypothesis with 90\% power at 5\% nominal $\alpha$ level is close to 9,000 per group. Faced with such a large average $\hat{N}$, the information carried in stage 1 data are eclipsed. This again  resembles a study with a single stage.

\begin{figure}[h]
\includegraphics[width=1\textwidth]{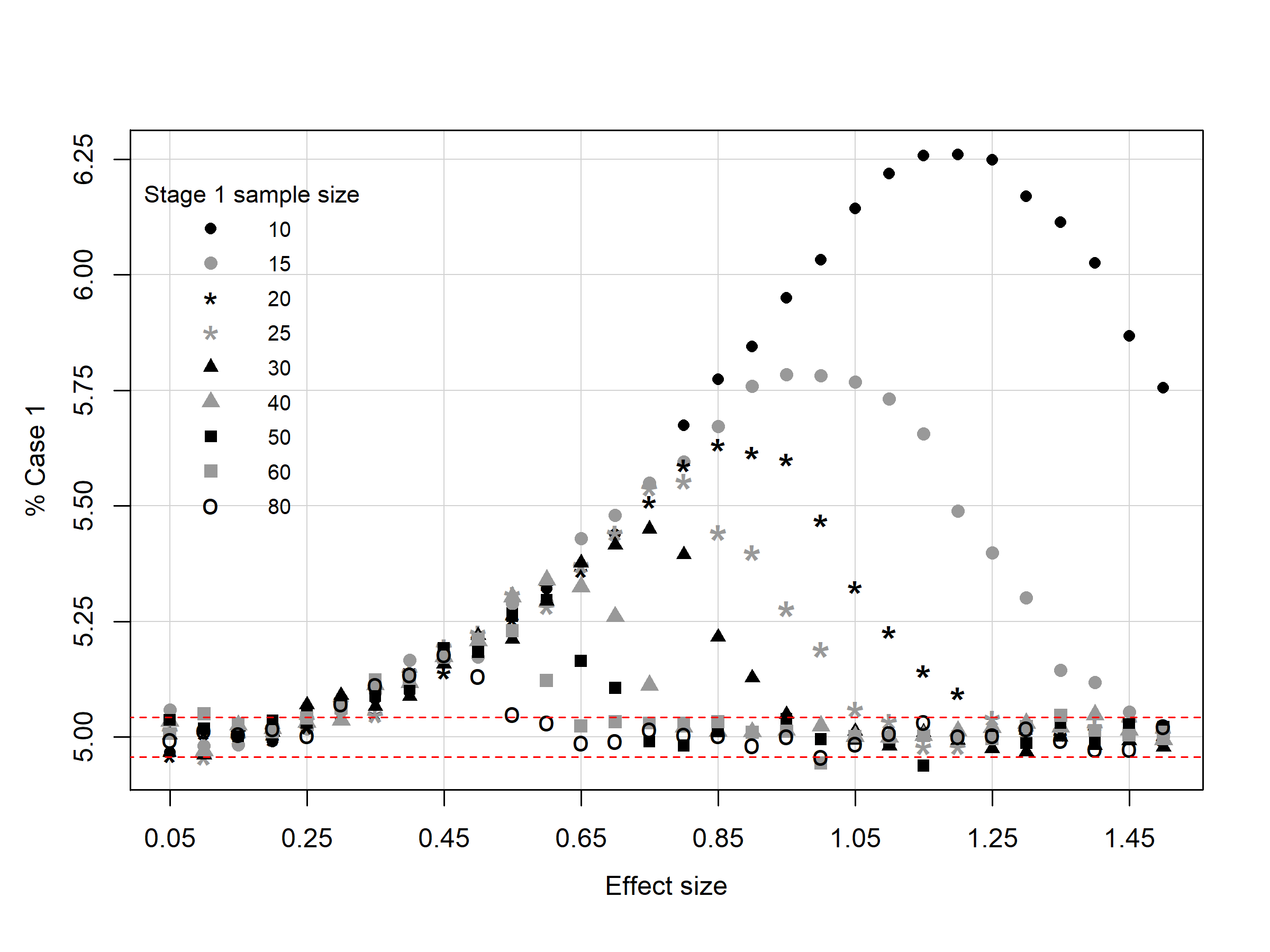}
\caption{$\alpha$ and inflation when $0\le m< \infty$.}\label{zero2inf}
\end{figure}

Figure \ref{zero2inf} shows the effect size ($x$-axis) vs. percent of Case 1, or type I error ($y$-axis) for an EQ test for all the $\tilde{n}$ we investigate, allowing $0 \le m < \infty$. At the small end of $\delta_0$ regardless of the size of $\tilde{n}$, $\alpha$ is around 5\% for the reason just stated.

\begin{figure}[h]
  \centering
  \includegraphics [width=1\textwidth]{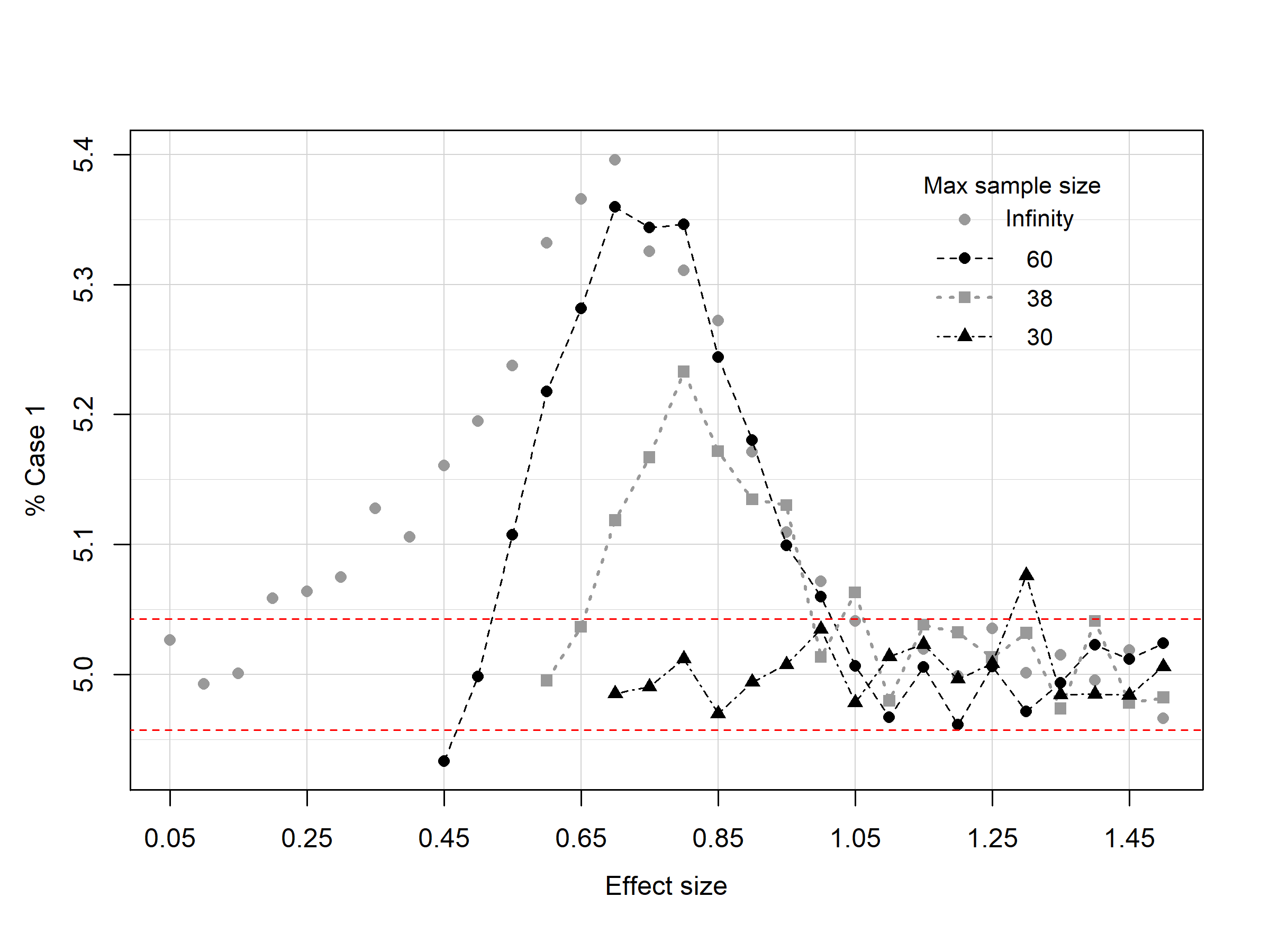} 
 \caption{Controlling \% Case 1 by decreasing $n_{Max}$. Figure example: $\tilde{n} = 15,\ n_{Min}=30$.}\label{nInt15Case1b}
\end{figure}

However, in reality the upper limit of the total sample size cannot be left un-capped. When $n_{Min}$ and $n_{Max}$ are predefined, since $\hat{N}$ is so large, $n_{Min}$ does not play any role, and $n_{Max}$ is always reached. What is more, since $n_{Max} << \infty$, there is almost no power to reject $H_{01}$, which is false. This implies Case 2 is the case predominating the false rejection of $H_{02}$. For example, Figure \ref{nInt15Case1b} gives results for $\tilde{n}=15$ and $n_{Min}$ fixed at 30 whereas $n_{Max}$ is infinity, 60, 38, or 30. 
As $n_{Max}$ decreases from 60 to 38 and then 30, the power for rejecting the false $H_{01}$ decreases, and Case 1\rq{}s become Case 2\rq{}s. Visually, the effect size where Case 1 rises to the range of 5\% shifts from 0.45, to 0.60 and 0.70, respectively. 

As the effect size moves away from 0, $\hat{N} < n_{Max}$ more and more often, and the variability in stage 2 sample size increases. Therefore, $\alpha$ starts to become more than its nominal level for the same reason given in Section \ref{sec_tilde_n_a}. 

At the other end of the spectrum where $\delta_0$ is large, Equation (\ref{eHatN}) gives us
$$
\lim_{\delta_0\rightarrow\infty} \hat{N} = \frac{\tilde{n}}{2\tilde{n}-1}\left(Z_{1-\beta/2}+Z_{1-\alpha}\right)^2,
$$
and
$$
\lim_{\delta_0\rightarrow\infty}SD(\hat{N}) = 0.
$$
With $\beta=0.10$ and $\alpha=0.05$, when $\tilde{n}=1$, $\lim_{\delta_0\rightarrow\infty}\hat{N} = 11$; when $\tilde{n}$ gets larger it converges to 5.5. 
In any practical setting where $\tilde{n}$ is at least 6, the stage 1 sample is already sufficient for the desired power. Therefore, any $n_{Min}\ge \tilde{n} \ge 6$ will have $n_{Min}$ as the final total sample size, and $n_{Max}$ will not play a role. This is why in Figure \ref{zero2inf} where $0\le m<\infty$, we can see a convergence toward 5\% for all $\tilde{n}$ at the larger end of $\delta_0$.

\begin{figure}[h]
  \centering
  \includegraphics [width=1\textwidth]{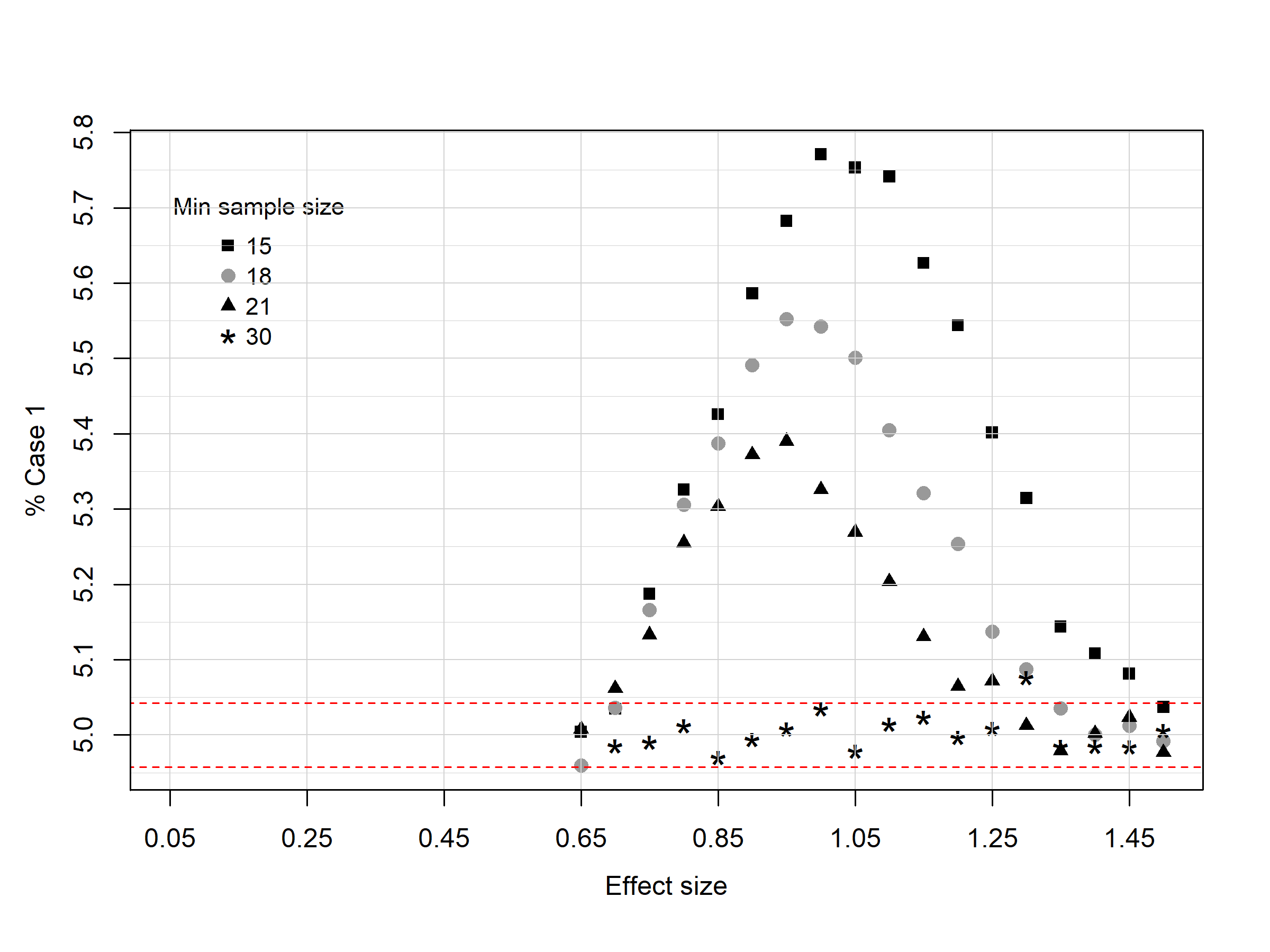}
 \caption{Controlling \% Case 1 by increasing the mandatory $n_{Min}$. Figure example: $\tilde{n} = 15,\ n_{Max}=30$.}\label{nInt15Case1c}
\end{figure}

Before $\delta_0$ becomes so large that $\hat{N}$ settles toward 5.5, $SD(\hat{N})$ is larger than 0, which can result in $\hat{N}$ that are above $n_{Min}$ but not equal to or exceeding $n_{Max}$. To reduce the variation in $\hat{N}$ and $\alpha$ inflation, we can set $n_{Min}$ just high enough to include all foreseeable cases where this lower limit of $\hat{N}$ suffices to reject $H_0$ with desired power at the nominal $\alpha$ level. From Figure \ref{nInt15Case1c} we can see that even with a moderate increase in the mandatory minimum total sample size, the $\alpha$ inflation reduces sharply.

\subsection{The peak $\alpha$ inflation}
As discussed in Section \ref{sec_typeI}, the precise type I error inflation depends on many factors such as stage 1 sample size, the equivalence margin, the sample size re-estimation rule which potentially includes upper or lower limits of the final sample size. It is difficult to give a general, analytical solution to the upper limit of the $\alpha$ inflation. As can be seen in all the figures, different versions of the SSR have led to different peaks both in terms of magnitude and the corresponding $\delta_0$ value even with the same stage 1 sample sizes.
Given the results from our simulation, if at the planning stage of a study, the range of assumed effect sizes falls on the smaller side of the effect size where peak $\alpha$ would occur, then it is important to impose a $n_{Max}$, since it can \lq\lq{}postpone\rq\rq{} the Case 1 from coming into the picture along the axis of effect size. It is clear though this solution results in having hardly any power for establishing the EQ or NI. Moreover, from a practical perspective, this is probably very rarely a relevant case. Because if it so happens that $\sigma^2$ is much larger in relation to $\delta_0$, it is simply hard to establish equivalence in the first place. In fact, in biosimilar pharmacokinetics studies where intra-individual variability is too large relative to the regulatory agencies\rq{} pre-defined equivalence range, FDA for example allows an approach where the reference range are scaled to the variability of the reference product \cite{Davit2012}.

On the other hand, if the assumed effect size for the study is on the larger side of where the peak $\alpha$ would occur, it is important to impose a larger $n_{Min}$ so it can force the convergence to the nominal $\alpha$ level earlier in terms of effect size. It is obvious that this solution can result in an overpowered study that is larger than necessary.

Table \ref{peakAlpha} shows the effect sizes and the peak $\alpha$ for each $\tilde{n}$ in our simulations where $0\le m < \infty$. They can serve as a reference for future planning. Although Table \ref{peakAlpha} refers to the type I error rate of an EQ test, for the NI test, the $\alpha$ are negligibly larger.

\begin{table}[h]
\caption{Peak $\alpha$ and its corresponding effect size \label{peakAlpha}}
\centering
\begin{tabular}{ccc}
\hline
$\tilde{n}$ & $\%$ Case 1& $\delta_0/\sigma$ \\ \hline
10 & 6.26 & 1.20  \\
15 & 5.78 & 0.95\\
20 & 5.63 & 0.85 \\
25 & 5.55 & 0.80 \\
30 & 5.45 & 0.75\\
40 & 5.34 & 0.60\\
50 & 5.30 & 0.60\\
60 & 5.23 & 0.55\\
80 & 5.18 & 0.45\\ \hline
\end{tabular}
\end{table}


\section{Summary}\label{sec_summ}

We have investigated causes of type I error inflation arising when blinded sample size reestimation is performed in equivalence trials. In Section \ref{sec_typeI}, we have given an explanation for why non-trivial type I error inflation can arise in this situation. The reasoning behind this explanation allows the construction of blinded sample size estimation rules which emphasize the inflation (one such rule is investigated in the Appendix). However, from an applied perspective, it is more important to understand how this phenomenon impacts sample size re-estimation rules which are used in practice.

To investigate this, we have performed extensive simulations using SSR outlined in Section \ref{ssr_rule}. We have set up the simulations to focus on scenarios where inflation must be suspected, but that are also practically relevant.

We find that the variation of the stage 2 sample size is a good, but not the only, indicator of (the magnitude of) type I error inflation. The stage 2 sample size is an increasing function of the total variance estimated at the interim. Hence, it is more variable when stage 1 is small and when there are no limits on the minimum and the maximum sample size allowed in stage 2. When the standardized EQ margin ($\delta_0/\sigma$) is very strict, stage 2 sample size is invariably estimated to be large in general and no inflation can arise. Conversely, when it is very generous, the stage 2 sample size tends to be always small and converges to a constant number as the margins tend towards infinity. Hence, there can also be no inflation. For this and practical reasons, the focus of our attention is on cases where the equivalence margin is roughly the same as the standard deviation.

We observe that if stage 2 sample size is not limited by additional restrictions, type I error inflations to $6.3\%$ can arise for stage 1 sample sizes of 10 subjects per group, a small, but not entirely unrealistic number, and persist ($5.2\%$) for as many as 80 subjects per group in stage 1.

These inflations are reduced when limits are placed on the allowed stage 2 sample sizes. In the range of practically relevant EQ margins, the lower limit ($n_{Min}$) is much more important than the upper limit ($n_{Max}$) here. The upper limit has an substantial impact on type I error probability only for values of the EQ margin that are a lot smaller than the standard deviation of the data.

Our investigation shows that the more similar to a fixed design a study is, the less $\alpha$ inflation can arise. This points to designing a study with a narrower range of $n_{Min}$ and $n_{Max}$. However, we also realize that when an interim is deemed necessary when planning a study, chances are there are uncertainties in key assumptions. Not having a lot of flexibility in choosing a second stage sample size limits the opportunities to rectify these uncertainties, hence, defeats the purpose of having an interim analysis and reduces the attractiveness of a blinded SSR. For an alternative, one can consider unblinded SSR for equivalence testing, see for example \cite{Maurer2018}.

As a recommendation for clinicians and data analysts who wish to incorporate a blinded SSR for a NI or EQ study, we suggest:
\begin{itemize}
\item[(1)] Choose a reasonably large interim sample size. From our simulations and recent experiences in planning clinical studies, $\tilde{n}$ should be at least 15.
\item[(2)] Allow $m\ge\tilde{n}$. In other words, impose a minimum cap $n_{Min}$ such that $n_{Min} \ge 2\tilde{n}$. If the stage 1 sample size is larger than 30, this can be relaxed.
\item[(3)] Limit the absolute maximum total. Our simulations show that $n_{Max}\le 3\tilde{n}$ provides reasonable $\alpha$ control.
\end{itemize}

When all these conditions are satisfied, maximum $\alpha$ can be limited to within 5.3\%. The third recommendation will probably be followed in practice already to avoid over costly big studies.

\appendix
\section{Numerical evaluation of type I error inflation for NI and EQ testing}\label{app_num}

Here we aim to demonstrate two points. First, we give a numerical example of a sample size re-estimation rule where an increased stage 1 sample size does not imply asymptotic decreasing of $\alpha$ inflation. Second, through this example we illustrate the do-ability of analytical computation of the exact $\alpha$ once the components of an SSR rule are defined. 

In this example, we assume 
that no additional subjects are recruited if the total variance estimate $\hat{\sigma}_T^2$, defined in Equation (\ref{totvar}), is less than or equal to a threshold $c^*$. When this happens, the study is unblinded, and a final hypothesis testing is conducted based on $d=\bar{y}_1-\bar{y}_2$, the observed difference between the mean responses in group 1 and group 2 after stage 1--the only stage, of the trial.

Since $
\hat{\sigma}_T^2=\frac{Q_1+Q_2}{n-1}$, we can express this as a boundary $c=(n-1)\cdot c^*$ on $Q_1+Q_2$.

For ease of notation, in this section, let stage 1 sample sizes for the two groups be  $n_1$ and $n_2$ such that $n_1=n_2$, and total stage 1 sample size $n=n_1+n_2=2n_1$. Therefore, $\sqrt{\frac{n_1}{2}}d\sim N(\sqrt{\frac{n_1}{2}}\delta,\ \sigma^2)$, and $Q_2=\frac{n_1}{2}d^2$. Consider the test
$$H_{02}:\delta\ge\delta_{up}$$ which will be rejected if
\[
t_{up}=\sqrt{\frac{n_1}{2}}\cdot\frac{d-\delta_{up}}{\sqrt{\frac{Q_1}{n-2}}}
\leq t_{\alpha}(n-2).
\]
It follows that $P(\mathrm{rejecting\ }H_{02}) = P\left(t_{up}\leq t_{\alpha}(n-2)\right)$ can be calculated in two steps: (1) the probability when $ Q_1+Q_2\leq c$, and (2) the probability when $ Q_1+Q_2 > c$.

As a numerical example, consider $\alpha=0.05$, $\sigma^2=1$, $\delta=\delta_{up}=0.5$. We look at the cases where $n_1=12,\ 24,\ 40$ and let  $c=n-1+\frac{n_1}{2}\delta^2$. We first calculate $P\left(t_{up}\leq t_{\alpha}(n-2)\ \big|\ Q_1+Q_2\leq c\right)$.

In general, for a given $Q_1=x$, we have
\begin{align*}
&P\left(t_{up}\leq t_{\alpha}(n-2)\ \big|\ Q_1=x\right)=&\\
&P\left(\sqrt{\frac{n_1}{2}}\cdot(d-\delta_{up})\leq q\cdot\sqrt{\frac{x}{n-2}}\ \right)=
\Phi\left(q\cdot\sqrt{\frac{x}{n-2}}-\sqrt{\frac{n_1}{2}}(\delta-\delta_{up})\right),
\end{align*}
where $\Phi(\cdot)$ denotes the cdf of $N(0,1)$, and $q=t_{\alpha}(n-2)$.

Conditional on a given $x$ with $x<c$, the joint distribution of rejecting $H_{02}$ and $Q_1+Q_2\le c$ is given by
\begin{align*}
 &P\left(t_{up}\leq t_{\alpha}(n-2),\ \sqrt{\frac{n_1}{2}} \left|d\right|\leq \sqrt{c-x}\ \bigg|\ Q_1=x\right)= &\\
 &P\left(
\sqrt{\frac{n_1}{2}}\cdot(d-\delta_{up})\leq q\cdot\sqrt{\frac{x}{n-2}},\
\sqrt{\frac{n_1}{2}}\cdot(d-\delta_{up})\leq \sqrt{c-x}-\sqrt{\frac{n_1}{2}}\delta_{up},\ \right.&\\
&\qquad\left.\sqrt{\frac{n_1}{2}}\cdot(d-\delta_{up})\geq -\sqrt{c-x}-\sqrt{\frac{n_1}{2}}\delta_{up}
\right)=&\\
&\Phi\left(\min\left(\sqrt{c-x}-\sqrt{\frac{n_1}{2}}\delta_{up},\ q\cdot\sqrt{\frac{x}{n-2}}\right)\right)-
\Phi\left(-\sqrt{c-x}-\sqrt{\frac{n_1}{2}}\delta_{up}\right)
\end{align*}
if $q\cdot\sqrt{\frac{x}{n-2}}>-\sqrt{c-x}-\sqrt{\frac{n_1}{2}}\delta_{up}$ and 0 otherwise.

Hence, the joint distribution of rejection of $H_{02}$ and $Q_1+Q_2\leq c$ is given by
\begin{align}
\label{A1}
 &P\left(t_{up}\leq t_{\alpha}(n-2),\ Q_1+Q_2\leq c\right)=&\nonumber\\
&\quad\int_{x=l^*}^{c^*} \bigg\{ \Phi\left(\min\left(\sqrt{c-x}-\sqrt{\frac{n_1}{2}}\delta_{up},\ q\cdot\sqrt{\frac{x}{n-2}}\ \right)\right)\nonumber\\
&\qquad -\Phi\left(-\sqrt{c-x}-\sqrt{\frac{n_1}{2}}\delta_{up}\right) \bigg\}\cdot f(x) dx,
\end{align}
where $f(\cdot)$ denotes the pdf of $\chi^2(n-2)$ and $0\leq l^*<c^*\leq c$ limit the range of integration in such a way that the integrand is larger than 0. The limits $l^*$ and $c^*$ are those solutions of the two equations
\begin{equation}\label{eqlim1}
q\cdot\sqrt{\frac{x}{n-2}}=-\sqrt{c-x}-\sqrt{\frac{n_1}{2}}\delta_{up}
\end{equation}
and
\begin{equation}\label{eqlim2}
q\cdot\sqrt{\frac{x}{n-2}}=\sqrt{c-x}-\sqrt{\frac{n_1}{2}}\delta_{up}
\end{equation}
which fall inside the interval $[0,c]$ (if there are such solutions). If no solution falls inside the interval $[0,c]$, then $l^*=0$ and $c^*=c$, respectively. Equations (\ref{eqlim1}) and (\ref{eqlim2}) can both be converted into quadratic equations and thus have closed-form solutions. To find a solution, we first have to check whether the solutions of the quadratic equations corresponding to (\ref{eqlim1}) or (\ref{eqlim2}) are actually solutions to (\ref{eqlim1}) or (\ref{eqlim2}), and then check which of these solutions fall into $[0,c]$. This is a mathematically straightforward, but somewhat tedious process.

The conditional distribution of rejection of $H_{02}$ given $Q_1+Q_2\leq c$ is thus
\[
P\left(t_{up}\leq t_{\alpha}(n-2),\ Q_1+Q_2\leq c\right)\cdot\left(P\left(Q_1+Q_2\leq c\right)\right)^{-1}.
\]

Formula (\ref{A1}) can be evaluated numerically (e.g. by SAS CALL QUAD). Likewise $P\left(Q_1+Q_2\leq c\right)$ is a probability from the non-central $\chi^2$-distribution $\chi^2(n-1;\frac{n_1}{2}\delta^2)$ that is implemented in most statistical software packages like SAS or R.

In the given numerical example, we have (calculations for $n_1=12$, numbers in brackets for $n_1=24,40$, respectively)
\begin{align*}
&P\left(t_{up}\leq t_{\alpha}(n-2), Q_1+Q_2\leq c\right)=0.0401 (0.0396,0.0393)&\\
&P\left(Q_1+Q_2\leq c\right)=0.5946 (0.5661,0.5510)&\\
&P\left(t_{up}\leq t_{\alpha}(n-2)\ \big|\ Q_1+Q_2\leq c\right)=\frac{0.0401}{0.5946}=0.0674 (0.0699,0.07125).
\end{align*}

Now we calculate the probability of rejecting $H_{02}$ when $Q_1+Q_2> c$. If the sample size in both groups is increased in such a way that stage 1 is rendered almost irrelevant. Then, $P\left(H_{02}\mbox{ rejected}\ \big|\  Q_1+Q_2>c\right)\approxeq 0.05$ and hence the type I error probability of this blinded SSR procedure at $\delta=\delta_{up}$ becomes
$0.5946\cdot 0.0674+(1-0.5946)\cdot 0.05=0.0603 (0.0612,0.0617)$ instead of the nominal $\alpha=0.05$.

Up until now, the discussion in this section focuses on NI testing when there is only one hypothesis, i.e. $H_{02}$, to be tested. Next we consider EQ testing where $H_{01}$ also need to be tested. Compare to NI testing, there is a decrease of the type I error which arises in the TOST strategy when $H_{02}$ is rejected but $H_{01}$ is not, i.e. Case 2 as defined in Section \ref{eqv_test}. This feature of TOST safe-guards the $\alpha$ inflation to some extend.

The joint probability that $Q_1+Q_2\leq c$, and both $H_{01}$ and $H_{02}$ are rejected is given by
\begin{align}\label{A2}
 &P\left(t_{low}\geq t_{1-\alpha}(n-2),\ t_{up}\leq t_{\alpha}(n-2),\ Q_1+Q_2\leq c\right)=&\\
 &P\left(Z\geq q^*(Q_1),\ Z\leq q\sqrt{\frac{Q_1}{n-2}},\ \right. \nonumber\\
 &\qquad \left. Z \in \left[-\sqrt{c-x}-\sqrt{\frac{n_1}{2}}\delta_{up},\ \sqrt{c-x}-\sqrt{\frac{n_1}{2}}\delta_{up}\right]\right)= \nonumber\\
 &\int_{x=l^*}^{c^*} \left(\Phi\left(\min\left(q\sqrt{\frac{x}{n-2}},\ \sqrt{c-x}-\sqrt{\frac{n_1}{2}}\delta_{up}\right)\right)- \right. \nonumber \\
 &\qquad \left. \Phi\left(\max\left(q^*(x),\ -\sqrt{c-x}-\sqrt{\frac{n_1}{2}}\delta_{up}\right)\right)\right)\cdot f(x) dx,\nonumber
 \end{align}
where $Z=\sqrt{\frac{n_1}{2}}(d-\delta_{up})\sim N(0,1)$, $q^*(x)=-q\sqrt{\frac{x}{n-2}}-\sqrt{\frac{n_1}{2}}(\delta_{up}-\delta_{low})$ and $0\leq l^*<c^*\leq c$ is such that over the range of $x\in \left[l^*,c^*\right]$,
\[
\min\left(q\sqrt{\frac{x}{n-2}},\ \sqrt{c-x}-\sqrt{\frac{n_1}{2}}\delta_{up}\right)>\max\left(q^*(x),\ -\sqrt{c-x}-\sqrt{\frac{n_1}{2}}\delta_{up}\right).
\]

For the example, let $\delta_{low}=-\delta_{up}$. Then we obtain (calculations for $n_1=12$, numbers in brackets for $n_1=24,40$, respectively):
\[
P\left(t_{low}\geq t_{1-\alpha}(n-2),\ t_{up}\leq t_{\alpha}(n-2),\ Q_1+Q_2\leq c\right)=0.0009\ (0.0193, 0.0379),
\]
and
\[
P\left(t_{low}\geq t_{1-\alpha}(n-2),\ t_{up}\leq t_{\alpha}(n-2)\ \big|\ Q_1+Q_2\leq c\right)
\]
is thus
\[
\frac{0.0009}{0.5946}=0.0015\ (0.0340, 0.0687).
\]
Finally, the unconditional rejection probability of both null hypotheses is
\[
P\left(\mbox{reject } H_{01} \mbox{ and } H_{02}\right)=0.0015+0.05\cdot(1-0.5946)=0.0212\ (0.0410, 0.0603).
\]


\end{document}